\documentclass[a4paper,11pt,nohyper]{JHEP3}
\def\d{\delta}

\def\t{\tau}

\def\S{\Sigma}

\def\del{\partial}              
   \let\d=\delta

 \let\t=\tau

\def\nn{\nonumber} \def\bd{\begin{document}} \def\ed{\end{document}}
\def\ds{\documentstyle} \let\fr=\frac \let\bl=\bigl \let\br=\bigr
\let\Br=\Bigr \let\Bl=\Bigl
\let\bm=\bibitem
\let\na=\nabla
\let\pa=\partial \let\ov=\overline
\newcommand{\be}{\begin{equation}}
\newcommand{\ee}{\end{equation}}
\def\ba{\begin{array}}
\def\ea{\end{array}}
\def\ft#1#2{{\textstyle{{\scriptstyle #1}\over {\scriptstyle #2}}}}
\def\fft#1#2{{#1 \over #2}}
\def\del{\partial}
\def\sst#1{{\scriptscriptstyle #1}}
 \def\oneone{\rlap 1\mkern4mu{\rm l}}
\def\ie{{\it i.e.\ }}
\def\via{{\it via}}
\def\semi{{\ltimes}}
\def\str{{\rm str}}
\def\Dm{{{D_{\sst{max}}}}}
\def\vac{ \left | 0 \right \rangle }
\def\kvac{ \left | k \right \rangle }

\def\sp{\; \; \;}

\def\bol{ \left | B (p^+) \right \rangle}
\def\bo1{ \left | B^0 (p^+) \right \rangle}

\def\bolt{ \left | B (p^+) \right \rangle_{\t}}

\def\boxl{ \left | B (x^-) \right \rangle}
\newcommand{\bea}{\begin{eqnarray}}
\newcommand{\eea}{\end{eqnarray}}

\def\<{ \langle }
\def\>{ \rangle }

\def\S{\Sigma}

\usepackage{amsmath,latexsym,amssymb,slashed}
\usepackage{graphicx}
\usepackage[latin1]{inputenc}

\renewcommand{\floatpagefraction}{0.6}
\renewcommand{\textfraction}{0.2}

\newcommand\ca{\mathcal{A}}
\newcommand\vp{\varphi}
\newcommand\beal{\begin{align}}
\newcommand\bbone{\ensuremath{\mathbbm{1}}}

\newcommand{\eq}[1]{\begin{equation}#1\end{equation}}
\newcommand{\spl}[1]{\begin{split}#1\end{split}}
\newcommand{\al}[1]{\begin{align}#1\end{align}}
\newcommand{\subeq}[1]{\begin{subequations}#1\end{subequations}}

\newcommand{\arXividhepth}[1]{\href{http://arxiv.org/abs/#1}arXiv:{\tt #1} [hep-th]}
\newcommand{\arXividother}[2]{\href{http://arxiv.org/abs/#1}arXiv:{\tt #1} [#2]}

\newcommand{\bg}[1]{\hat{#1}}
\newcommand{\wj}{\widetilde{J}}
\newcommand{\reo}{\mathrm{Re}~\!\omega}
\newcommand{\imo}{\mathrm{Im}~\!\omega}
\newcommand{\ads}{AdS_4}
\newcommand{\mcal}{\mathcal{M}}
\newcommand{\ccal}{\mathcal{C}}
\newcommand{\ncal}{\mathcal{N}}
\newcommand{\boxedeq}[1]{
\begin{equation}
\fbox{
\rule[0.7cm]{0pt}{0pt}
$#1$
\rule[-0.45cm]{0pt}{0pt}
}
\end{equation}
}

\def\d{\text{d}}

\def\slashchar#1{\setbox0=\hbox{$#1$}           
\dimen0=\wd0                                 
\setbox1=\hbox{/} \dimen1=\wd1               
\ifdim\dimen0>\dimen1                        
\rlap{\hbox to \dimen0{\hfil/\hfil}}      
#1                                        
\else                                        
\rlap{\hbox to \dimen1{\hfil$#1$\hfil}}   
/                                         
\fi}

\def\Re           {{\rm Re\hskip0.1em}}
\def\Im           {{\rm Im\hskip0.1em}}

\newcommand{\E}{\text{\tiny E}}

\title{A consistent truncation of IIB supergravity on manifolds admitting a
Sasaki-Einstein structure}

\author{Kostas Skenderis${}^{\diamondsuit, \spadesuit}$, Marika Taylor${}^{\diamondsuit}$  and  Dimitrios Tsimpis${}^{\clubsuit}$  \\

\begin{itemize}
  \renewcommand{\labelitemi}{${}^\diamondsuit$}
\item  Institute of Theoretical Physics \\
Valckenierstraat 65, 1018 XE Amsterdam \\
The Netherlands
\renewcommand{\labelitemi}{${}^\spadesuit$}
\item Korteweg-de Vries Institute for Mathematics \\
Science Park 904, 1098 XH Amsterdam \\
The Netherlands
  \renewcommand{\labelitemi}{${}^\clubsuit$}
\item  Arnold-Sommerfeld-Center for Theoretical Physics\\
Department f\"ur Physik, Ludwig-Maximilians-Universit\"at M\"unchen\\
Theresienstra\ss e 37, 80333 M\"unchen, Germany
  \end{itemize}

\bigskip
 E-mail:
 \email{{k.skenderis@uva.nl}, \email{m.taylor@uva.nl}, \email{dimitrios.tsimpis@lmu.de}}}

\abstract{We present a consistent truncation of IIB supergravity on
manifolds admitting a Sasaki-Einstein structure,
which keeps the metric and five real scalar fields. This theory can be further
truncated to a constrained one-parameter family that depends on only the metric and one scalar, as well as to a theory with a metric and three scalars.
The reduced theory admits supersymmetric and
non-supersymmetric $AdS_5$ and $AdS_4 \times \mathbb{R}$ solutions.
We analyze the spectrum around the $AdS$ critical points and identify the dual operators.}

\begin{document}

 \newcommand{\bc}{\begin{center}}
 \newcommand{\ec}{\end{center}}
 \newcommand{\bfr}{\begin{flushright}}
 \newcommand{\efr}{\end{flushright}}
 \newcommand{\bfl}{\begin{flushleft}}
 \newcommand{\efl}{\end{flushleft}}
 \newcommand{\bt}{\begin{tabular}}
 \newcommand{\et}{\end{tabular}}

\section{Introduction}

Consistent truncation of supergravity theories on compact manifolds has proven to
be a useful tool in holography. The truncation reduces the ten-dimensional
equations of motion to simpler five-dimensional equations, which in turn makes it easier to find
explicit solutions and facilitates the construction of a precise holographic dictionary
by making tractable the systematic analysis of the asymptotic structure of
the solutions. Moreover, explicit reduction formulae can be used to uplift
all results to ten dimensions and discuss their physical content both from the five-
and  ten-dimensional perspective.

Consistent truncation is not a prerequisite for discussing the system from a
five-dimensional perspective and performing a holographic analysis.
This can be done without consistent truncation, using the
framework of Kaluza-Klein holography \cite{Skenderis:2006uy}
which organizes the reduction of all modes according to the way they
contribute to the 1-point functions of dual operators:
in the computation of the 1-point function of an operator of dimension $\Delta$ only
a finite number of fields can contribute, namely the ones
that are dual to operators with dimension less than $\Delta$.
Thus, effectively one deals with finite number of fields (whose number depends on the
dimension of the operator under consideration).
In contrast, in consistent truncation one keeps a finite number of fields from the outset.
The fact that this can be done implies that in the dual theory there is a subset of
operators that close under OPEs (at least, in the large $N$, strong coupling limit).

Consistent truncations of maximal supergravities
over spheres have been  studied extensively \cite{spheres}, and it is generally believed that the five-dimensional $\mathcal{N}=8$ supergravity is a consistent
truncation of ten-dimensional IIB supergravity on the five-sphere.
Spheres are rather special cases of Sasaki-Einstein manifolds, and
one of the motivations of the present work was to furnish consistent
truncations of IIB on more general five-dimensional Sasaki-Einstein spaces.
 More recently, two consistent non-supersymmetric truncations of IIB
supergravity on five-dimensional Sasaki-Einstein manifolds were presented in
\cite{mmt}; we will return to the precise relation of our results to those
of \cite{mmt} in the following. For other consistent truncations see \cite{gaunt}.

Another motivation of the present work
was to understand holography for the recently found solutions of \cite{lt}. These are solutions of the form
$AdS_4 \times \mathbb{R} \times \mathcal{M}_5$, where  $\mathcal{M}_5$
is a squashed Sasaki-Einstein five-dimensional manifold. These solutions were
constructed by exploiting a general geometrical property of all such manifolds,
namely their local $SU(2)$ structure, and are therefore independent of the details of $\mathcal{M}_5$. In particular, $\mathcal{M}_5$ need not be a five-sphere, but can be any five-dimensional manifold admitting a Sasaki-Einstein structure.

It is natural to ask whether the solutions of
\cite{lt} can be understood within the framework of an effective
theory in five dimensions, as is the case for the non-supersymmetric
$AdS_4 \times \mathbb{R} \times {SE}_5$ solution of \cite{fake}.\footnote{
It is straightforward to generalize the solution of  \cite{fake} so that
the five-dimensional internal manifold is a general Sasaki-Einstein space.}
Given the observations of the preceding paragraph, such an effective theory
should be a consistent truncation of IIB on $\mathcal{M}_5$, where $\mathcal{M}_5$ is any five-dimensional manifold admitting a local $SU(2)$ structure of Sasaki-Einstein type.
As we will see in the following, the
solutions of both \cite{lt} and \cite{fake}, the supersymmetric $AdS_5\times SE_5$ solutions, as well as
the non-supersymmetric  $AdS_5\times \widetilde{SE}_5$ solutions of \cite{romans} (where
$\widetilde{SE}_5$ is squashed  Sasaki-Einstein)
can indeed all be recovered as solutions of a five-dimensional consistent truncation of IIB supergravity.

The consistent truncation mentioned in the preceding paragraph is
 the main result of the present paper, and is presented in eq.~(\ref{leff})
below. It is a five-dimensional non-supersymmetric (bosonic) theory consisting of gravity coupled to five real scalars. Moreover, it is possible to show that the theory admits two different further truncations. On the one hand,
it reduces to the gravity-scalar
sector of the three-scalar truncation of \cite{mmt} upon suitably eliminating
two of the five scalars. From the ten-dimensional point of view,
this operation  corresponds to setting the three-form flux to zero. On the other hand, the theory can be truncated to a constrained Lagrangian containing the graviton and one real scalar.

Our effective five-dimensional theory admits two $AdS_5$ critical points, one supersymmetric and one non-supersymmetric. These
uplift to IIB vacua of the form $AdS_5\times \mathcal{M}_5$, where $\mathcal{M}_5$ is a round, squashed $SE_5$ in the supersymmetric, non-supersymmetric  case respectively. Specializing to the case where
the $SE_5$ is the five-sphere allows us to unambiguously identify the dual field-theory operators, at least in the supersymmetric case, using the
well-established AdS/CFT dictionary. However, the fact that none of the fields in our truncation corresponds to higher harmonics on $\mathcal{M}_5$ suggests that our analysis may have more general validity.

The remainder of the paper is organized as follows. The Lagrangian of the  five-dimensional consistent truncation is presented in section \ref{sec1}. The three-scalar sector of \cite{mmt} is recovered in section \ref{sec:three} as
a further truncation.
The truncation to the constrained single-scalar theory is described
in section \ref{family}. The four classes of supersymmetric and
non-supersymmetric $AdS_5$ and $AdS_4\times \mathbb{R}$ solutions
are presented in sections \ref{sec:3.1} to \ref{sec:3.4}. The
masses of the fluctuations around the two critical $AdS_5$ points
and the spectrum of dual operators are discussed in section \ref{masses}. We
conclude with a discussion of  future directions in section \ref{conc}.
Appendix \ref{app:se} contains some relevant facts about five-dimensional Sasaki-Einstein geometry, while appendix \ref{app:truncation} contains
the technical details of the consistent truncation.

\section{The five-dimensional effective action}\label{sec1}

We will now present a consistent truncation of the ten-dimensional
IIB supergravity to a five-dimensional effective action containing
the graviton and five real scalars.
In order to avoid making this section overly technical we will
limit ourselves to a schematic description of the
reduction ansatz, deferring all technical details
 to appendix \ref{app:truncation}.

The reduction ansatz for the 10D
metric in the Einstein frame reads:
\eq{\label{g10e}
ds^2_{10}=g_{\mu\nu}dx^\mu dx^\nu + ds^2(\mathcal{M}_5)
~,}
where ($\mu$,$\nu$) are indices along the 5D spacetime, and $\mathcal{M}_5$ is the five-dimensional internal manifold. The latter
can be obtained from an associated `unit' Sasaki-Einstein metric
$ds_{SE}^2(\mathcal{M}_5)$ through
a certain deformation. This metric deformation is parameterized
by two real scalars  $\psi$, $\omega$, and corresponds to
an overall warping together with a squashing (also sometimes called
`stretching' -- depending on one's outlook) of the $U(1)$ fiber of the
Sasaki-Einstein space. Schematically we have:
\be\label{5dmetricschematic}
ds^2(\mathcal{M}_5)\sim~\! e^{2\psi(x)}ds^2_{KE}
+~\!e^{2\omega(x)}u\otimes u~,
\ee
where
\be
ds^2_{KE}+ u\otimes u~,
\ee
is the metric of a canonically normalized `unit' five-dimensional
Sasaki-Einstein space ({\it cf.{}} appendix \ref{app:se}  for the relevant  facts about five-dimensional Sasaki-Einstein manifolds). The precise expression for the internal metric is given in eq.~(\ref{5dmetric}).
The two scalars $\psi$, $\omega$ are assumed to depend only on the 5D spacetime coordinates $x^{\mu}$.

The dilaton $\phi$ is assumed to be an independent 5D spacetime scalar. In addition the ansatz includes two more real 5D spacetime scalars, for a total of
five real scalars: a scalar $\chi$
which parameterizes the NSNS three-form flux
\be\label{nsfluxschematic}
H\sim~\!  \mathrm{Re}\Omega\wedge d\chi
+~\! {\chi}J\wedge u
~,
\ee
and a scalar $\varphi$
which enters in the ansatz for the RR three-form flux
\eq{
 F_3\sim~\!  J\wedge d\varphi+~\! {\varphi}\mathrm{Re}\Omega\wedge u
~.}
The  real two-form $J$, the complex two-form $\Omega$ and the real one-form $u$ are related to the underlying Sasaki-Einstein structure of the internal manifold. Moreover, the RR one-form flux is assumed to vanish,
\eq{
F_1=0
~,}
while the RR five-form flux depends, in addition to
the dilaton, on both $\chi$, $\varphi$; it consists of
two components: one which is
proportional to the volume of $\mathcal{M}_5$,
\eq{
F_5\sim ~\!  J\wedge J\wedge u
~,}
and another along the 5D spacetime  directions which
is obtained from the
above by ten-dimensional Hodge-dualization, so that the
total five-form flux is self-dual.
The precise expressions
of the ans\"{a}tze for the fluxes are given in eqs.~(\ref{nsflux},\ref{rrflux}).

As shown in detail in appendix \ref{app:truncation}, this reduction ansatz
leads to a consistent truncation of ten-dimensional
IIB supergravity. The resulting  five-dimensional effective action is given by:
\boxedeq{\spl{\label{leff}
\mathcal{L}_{eff}&=e^{4\psi+\omega}\left\{
R^{(5)}+12(\partial\psi)^2+8\partial\psi\cdot\partial\omega-\frac{1}{2}(\partial \phi)^2\right\}\\
&-\frac{1}{5}e^\omega\left\{
e^{\phi}(\partial\varphi)^2
+e^{-\phi}(\partial\chi)^2
\right\}\\
&-2W^2(2-\frac{1}{2}{\chi\varphi})^2e^{-4\psi-\omega}+
20W^2\left(1-\frac{e^{2\omega}}{5e^{2\psi}}
\right)e^{2\psi+\omega}\\
&-\frac{5}{4}W^2e^{-\omega}\left(
e^{\phi}\varphi^2
+e^{-\phi}\chi^2
\right)
~,}}
where $W$ is a real constant.

\subsection{A three-scalar truncation}\label{sec:three}

The effective action (\ref{leff}) can be further consistently truncated to three scalars, in the limit where the scalars $\varphi$, $\chi$ are eliminated:
\eq{\label{ctr}
{\varphi},~{\chi}\rightarrow 0~.
}
As can be seen from eqs.~(\ref{nsflux},\ref{rrflux}), this corresponds from the
ten-dimensional point of view to setting the three-form flux to zero. It can be easily verified that this limit is consistent with the  $\varphi$, $\chi$ equations of motion (\ref{r5},\ref{r6}).

In \cite{mmt}, a consistent non-supersymmetric IIB truncation was presented which corresponds
to vanishing ten-dimensional three-form flux and whose scalar sector contains
precisely three scalars.
It can be seen that the further truncation (\ref{ctr}) of the effective action (\ref{leff}) is none other than the gravity-scalar sector of the ($m^2=8$) consistent
truncation of \cite{mmt}.

Indeed, this can be explicitly seen as follows. Starting with the effective action (\ref{leff}) we can pass to the Einstein frame by setting:
\eq{
g_{\mu\nu}=e^{-\frac{2}{3}(4\psi+\omega)}g^{E}_{\mu\nu}~.
}
Moreover, we can diagonalize the resulting kinetic terms by setting:
\eq{
\psi=\frac14(2u-3v)~;~~~~~\omega=-\frac14(8u-3v)~,
}
upon which the effective action reads:
\eq{\spl{
\mathcal{L}_{eff}&=
R_E^{(5)}-5(\partial u)^2-\frac{15}{2}(\partial v)^2-\frac{1}{2}(\partial \phi)^2
+W^2\left\{
-8e^{-10v}+20e^{-u-4v}-4e^{-6u-4v}
\right\}
~.}}
The above can be made to match precisely with the gravity-scalar sector of eq.(4.10) of ref.~\cite{mmt} upon performing the constant shifts:
\eq{
u\rightarrow u+\frac15\log \frac65~;~~~~~v\rightarrow v+\frac15\log \frac65~,
}
and setting $W={6}/{5}$.

\section{A one-parameter family of constrained Lagrangians}\label{family}

The effective Lagrangian (\ref{leff}) admits a one-parameter family of
consistent truncations to a single scalar, which we now describe in detail.

The truncation is obtained by setting:
\eq{\spl{\label{alphdef}
\varphi&=\alpha~\!e^{-\phi/2}\\
\chi&=\alpha~\!e^{\phi/2}
}}
and
\eq{\label{65}
\alpha,~\psi,~\omega=\mathrm{constant}
~.
}
With the above assumptions the dilaton equation (\ref{r1})
reduces to:
\eq{\label{72}
\nabla^2\phi=0~,
}
while the three-form equations (\ref{r5},\ref{r6}) both reduce to:
\boxedeq{\label{effconstr}
\alpha\left\{(\partial \phi)^2-20W^2x^{-1}\left(
(2-\frac{1}{2}{\alpha}^2)y^{-2}-\frac{5}{4}
\right)\right\}=0~,
}
where we set $y:=e^{2\psi}$, $x:=e^{2\omega}$.
The internal part of Einstein equations (\ref{r2},\ref{r3}) reduces to:
\eq{\spl{\label{sys2}
2x^2+5xy&=2(2-\frac{1}{2}{\alpha}^2)^2y^{-2}+\frac{5}{2}{\alpha}^2\\
8x^2&=(2-\frac{1}{2}{\alpha}^2)(2-{\alpha}^2)y^{-2}+5{\alpha}^2
~.}}
The system of equations (\ref{sys2}) determines  $x$ (equiv.~$\omega$), $y$ (equiv.~$\psi$) in terms of the free parameter $\alpha$.
Finally, the external Einstein equations reduce to:
\eq{\label{78}
R_{\mu\nu}=\frac{1}{2}(1+\frac{\alpha^2}{5y^{2}})\partial_\mu \phi\partial_\nu \phi
-\frac{1}{3}g_{\mu\nu}\mu(\alpha)
~,
}
where we defined
\eq{
\mu(\alpha):=\frac{20W^2}{y}\left(1-\frac{x}{5y} \right)
-\frac{2W^2}{xy^4}(2-\frac{1}{2}{\alpha}^2)^2-\frac{5W^2}{2xy^2}{\alpha}^2
~.}
The equations above can be integrated to an effective action
\boxedeq{\label{leffprimeprime}
\mathcal{L}^{\prime}_{eff}
=R^{(5)}-\frac{1}{2}\left(1+\frac{{\alpha}^2}{5y^{2}}\right)(\partial \phi)^2+\mu(\alpha)~,
}
{\it subject to the constraint} (\ref{effconstr}).

The system (\ref{effconstr},\ref{leffprimeprime}) admits a number of supersymmetric and non-supersymmetric solutions for different
values of the free constant parameter $\alpha$, to which we turn in the
following.

\subsection{Supersymmetric $AdS_5\times SE_5$}\label{sec:3.1}

The supersymmetric $AdS_5\times SE_5$ solution (with
maximal supersymmetry in the case where $SE_5$ is an $S^5$) can be obtained
from (\ref{leffprimeprime},\ref{effconstr}) by setting  the dilaton $\phi$ to a constant:
\eq{
\phi=\phi_0~,
}
and in addition taking the limit
\eq{
\alpha\rightarrow 0~.
}
Eqs.~(\ref{72},\ref{effconstr}) are then automatically satisfied, while
(\ref{alphdef}) implies:
\eq{\label{ansa2}
{\varphi}={\chi}=0~.
}
The system (\ref{sys2}) is then solved for
\eq{\label{ansa1}
c_1e^{2\psi}=c_2e^{2\omega}=\frac{1}{\lambda^2}
~,}
where we have set:
\eq{\label{54}
\lambda^2:=\left(\frac{5}{6}\right)^{\frac{3}{2}}W^2 ~.
}
From the ten-dimensional point of view, eqs.~(\ref{ansa1},\ref{5dmetric}) imply that
the internal space is  Sasaki-Einstein:
\eq{
ds^2(\mathcal{M}_5)=\frac{1}{\lambda^2}\left(
ds^2_{KE}+u\otimes u
\right)
~,
}
{\it cf.} appendix \ref{app:se}. Moreover, as can be seen from (\ref{nsflux},\ref{rrflux}), eq.~(\ref{ansa2}) sets all fluxes to zero except for the
RR five-form:
\eq{
F_5=4\lambda vol_{5}
~,}
where $vol_{5}$ denotes the volume of the
internal space.
Finally, the Einstein eqs.(\ref{78}) reduce to:
\eq{
R_{\mu\nu}=-4\lambda^2g_{\mu\nu}
~,}
i.e. the external space is $AdS_5$ with inverse radius $\lambda$.

\subsection{Non-supersymmetric $AdS_5\times \widetilde{SE}_5$}\label{sec:3.2}

A non-supersymmetric $AdS_5\times \mathcal{M}_5$ solution can be obtained
from (\ref{leffprimeprime},\ref{effconstr}) by setting  the dilaton $\phi$ to a constant:
\eq{
\phi=\phi_0~,
}
and in addition setting:
\eq{
{\alpha}^2=2~;~~~~~
e^{2\psi}=\frac{2}{\sqrt{5}}~;~~~~~
e^{2\omega}=\frac{\sqrt{5}}{2}
~.
}
As is straightforward to verify, the above assignments solve eqs.~(\ref{72},\ref{effconstr},\ref{sys2}).
From the ten-dimensional point of view
the internal space $\mathcal{M}_5$ is a squashed five-dimensional  Sasaki-Einstein, as
can be seen from (\ref{5dmetric}).

Finally, the Einstein eqs.(\ref{78}) reduce to:
\eq{
R_{\mu\nu}=-\frac{5\sqrt{5}}{4}W^2g_{\mu\nu}
~,}
i.e. the external space is $AdS_5$ with inverse radius $\lambda$, where:
\eq{\label{61}
\lambda^2=\frac{5\sqrt{5}}{16}W^2
~.}
This non-supersymmetric $AdS_5$ solution was found in \cite{romans}, and its spectrum was discussed in \cite{Distler:1998gb}. In \cite{Freedman:1999gp} a truncation of ${\cal N}$=8 gauged supergravity to
the graviton and four scalars was considered. This truncation admits both the maximally supersymmetric $AdS_5$ vacuum
and a non-supersymmetric $AdS_5$ critical point. The masses of the four scalars at the maximally supersymmetric critical point are such that two of the dual operators are of dimension two, whilst the other two are of dimension three and four. As we will see shortly, the latter two operators are retained in our consistent truncation also. In the non-supersymmetric $AdS_5$ critical point of \cite{Freedman:1999gp}, it is the scalar field dual to the operator which is of dimension three in the supersymmetric vacuum which acquires a finite value. In our non-supersymmetric $AdS_5$ vacuum, finite values are acquired by
both this scalar field and other scalar fields not retained in the truncated action used in \cite{Freedman:1999gp}. It would be interesting to explore how these critical points are related to each other.

\subsection{Supersymmetric $AdS_4\times\mathbb{R}\times\widetilde{SE}_5$}\label{sec:3.3}

This  solution uplifts to the supersymmetric 10D IIB solution
presented in \cite{lt};\footnote{Note however that we will only be able to make contact with those solutions of \cite{lt} for which $F_1=0$.} it can be obtained as a solution of the constrained effective Lagrangian (\ref{leffprimeprime}) by setting:
\eq{
\alpha=1~,~~\omega=\psi=0
~.}
The above implies that from the ten-dimensional point of view
the internal space $\mathcal{M}_5$ is
a squashed Sasaki-Einstein, as can be seen from  eq.(\ref{5dmetric}).
Moreover the scalars $\phi$, $\varphi$, $\chi$ are assumed to only depend
on a single  coordinate $r$, and:
\eq{
\frac{1}{4}\phi^\prime=-\frac{\varphi^\prime}{2\varphi}
=\frac{\chi^\prime}{2\chi}=\beta
~,
}
where the prime denotes differentiation with respect to $r$ and $\beta$ is a constant. In other words the scalars  $\phi$, $\log\varphi$, $\log\chi$ depend linearly on $r$.

With these substitutions, eqs.~(\ref{72},\ref{sys2}) are satisfied, while eq.~(\ref{effconstr}) gives:
\eq{\label{betavalue}
\beta=\frac{\sqrt{5}}{4}W
~.}
Finally, the Einstein eqs.~(\ref{78}) are solved for a geometry of the form $AdS_4\times\mathbb{R}$:
\eq{
ds^2=ds^2(AdS_4)+dr^2~.
}
More precisely,
the components of the Einstein eqs.~(\ref{78}) along the
$AdS_4$ directions reduce to:
\eq{
R^{(4)}_{\mu\nu}=-3W^2g^{(4)}_{\mu\nu}
~,}
which is indeed satisfied by an $AdS_4$ space of inverse radius $W$;
 the $rr$ component of (\ref{78}) is
satisfied by virtue of (\ref{betavalue}), while the
the mixed components are satisfied automatically.

In \cite{fake} $AdS_D$-sliced domain walls were studied within the
context of `fake supergravity', using an adapted superpotential
as a useful analysis tool.
In order to make contact with their formalism, one must set $\phi=h=0$, $d=4$ in (2.11) of \cite{fake} and moreover identify (there$\rightarrow$here):
\eq{\spl{
\kappa\phi^\prime&\rightarrow \sqrt{\frac{3}{5}}\phi^\prime\\
\kappa^2V&\rightarrow -\frac{9}{2}W^2\\
L_4^2&\rightarrow \frac{1}{W^2}~.
}}
In particular, the (non-perturbative) stability results established in \cite{fake}
carry over. Furthermore, the Janus solution of \cite{Bak:2003jk} is also a solution of our
effective action and thus can be lifted to 10 dimensions using an arbitrary five dimensional
manifold admitting a Sasaki-Einstein structure.

\subsection{Non-supersymmetric $AdS_4\times\mathbb{R}\times{SE}_5$}\label{sec:3.4}

This is the non-supersymmetric solution presented in \cite{fake}.
This solution was first found in \cite{Robb:1984uj,Quevedo:1985zx}.
It can be obtained as a solution of the constrained effective Lagrangian
(\ref{leffprimeprime}) in the limit
\eq{
\alpha\rightarrow 0~.
}
As in the supersymmetric solution the constraint (\ref{effconstr}) is then automatically satisfied, while
(\ref{alphdef}) implies:
\eq{\label{ansa2b}
{\varphi}={\chi}=0~.
}
The system (\ref{sys2}) is then solved for
\eq{\label{ansa1b}
c_1e^{2\psi}=c_2e^{2\omega}=\frac{1}{\lambda^2}
~,}
where we have set:
\eq{
\lambda^2:=\left(\frac{5}{6}\right)^{\frac{3}{2}}W^2 ~.
}
As in the  supersymmetric case, the equations above  imply that
the internal space is Sasaki-Einstein and  all fluxes are zero except for the
RR five-form:
\eq{
F_5=4\lambda vol_{5}
~,}
where $ vol_{5}$ denotes the volume form of the internal
 space. Moreover the scalars $\phi$, $\varphi$, $\chi$ are assumed to only depend
on a single  coordinate $r$, and:
\eq{
\frac{1}{4}\phi^\prime=-\frac{\varphi^\prime}{2\varphi}
=\frac{\chi^\prime}{2\chi}=\beta
~,
}
where the prime denotes differentiation with respect to $r$ and $\beta$ is a constant. In other words the scalars  $\phi$, $\log\varphi$, $\log\chi$  depend linearly on $r$.

With these substitutions eqs.~(\ref{72},\ref{effconstr},\ref{sys2}) are satisfied. Moreover, the Einstein eqs.~(\ref{78}) are solved for a geometry of the form $AdS_4\times\mathbb{R}$:
\eq{
ds^2=ds^2(AdS_4)+dr^2~.
}
More precisely,
the components of the Einstein eqn.~(\ref{78}) along the
$AdS_4$ directions reduce to:
\eq{
R^{(4)}_{\mu\nu}=-4\lambda^2g^{(4)}_{\mu\nu}
~,}
which is indeed satisfied by an $AdS_4$ space of inverse square radius $4\lambda^2/3$;
 the $rr$ component of (\ref{78}) is
satisfied provided:
\eq{
\beta^2=\frac{1}{2}\lambda^2
~,}
while the mixed components are satisfied automatically.

To translate between \cite{fake} and the present, one must set $\phi=h=0$, $d=4$ in (2.11) of \cite{fake} and moreover identify (there$\rightarrow$here):
\eq{\spl{
\kappa\phi^\prime&\rightarrow \frac{1}{\sqrt{2}}\phi^\prime\\
\kappa^2V&\rightarrow -6\lambda^2\\
L_4^2&\rightarrow \frac{3}{4\lambda^2}~.
}}
One can then use the results of this
subsection to find a different uplift of the Janus solution.
We can also adopt the results of \cite{fake} on stability.

\section{Masses}\label{masses}

We will now obtain the masses of the fluctuations around the two
critical $AdS_5\times \mathcal{M}_5$ solutions described earlier.
Although the spectra we obtain are valid for general  $\mathcal{M}_5$,
as long as it admits a Sasaki-Einstein structure,
in the supersymmetric case of section \ref{susymass}
it will be useful to specialize to the
case where $\mathcal{M}_5$ is the five-sphere, in order to identify
the dual CFT$_4$ operators using the well-known AdS/CFT dictionary.

In both cases we will take $\lambda$ to be the inverse radius of $AdS_5$, i.e:
\eq{
R=-20\lambda^2
~,}
where $R$ is the scalar curvature of $AdS_5$.
Our conventions are such that for a scalar  particle $\sigma$ with equation of motion:
\eq{
(\nabla^2-M_\sigma^2)~\!\sigma=0~,
}
the Breitenlohner-Freedman bound in five dimensions reads:
\eq{\label{bf}
M^2\geq-4\lambda^2~.
}

\subsection{The  supersymmetric $AdS_5\times \mathcal{M}_5$}\label{susymass}

Let us expand all fields around the critical point as follows:
\eq{\spl{
\phi&=\phi_0+\delta \phi\\
e^{2\psi}&=\frac{1}{c_1}\left( \frac{1}{\lambda^2}+\delta\psi \right) \\
e^{2\omega}&=\frac{1}{c_2}\left( \frac{1}{\lambda^2}+\delta\omega \right) \\
{\varphi}&=e^{-\phi_0/2}\delta\varphi\\
{\chi}&=e^{\phi_0/2}\delta\chi
~,}}
where $\phi_0$ is a constant. The inverse radius $\lambda$ of $AdS_5$ is related to the constant $W$ through eq.~(\ref{54}).
%
%

Expanding the equations of motion
to first order in the fluctuations we find the following mass spectrum:

\bigskip

\begin{center}

\begin{tabular}{| c || c |c|c|c|}
\hline
$\sigma$ &  $(M_\sigma/\lambda)^2$  & 10D & $\mathcal{O}$ & $\Delta$\\
\hline
\hline
$\delta \phi$  &  0   & $B$& tr$F_+^2$    &4\\
\hline
$4\delta\psi+\delta\omega$ &  32  & $h_{\alpha}^\alpha$ &  tr$F_+^2 F_-^2$    &8\\
\hline
$\delta\psi-\delta\omega$  &  12  & ${h}_{(\alpha\beta)}$ & tr$\lambda\lambda\bar{\lambda}\bar{\lambda}$   &6\\
\hline
$ \delta\varphi+\delta\chi$ &   $ -3$ & ${A}_{\alpha\beta}$ & tr$\lambda\lambda$  &3\\
\hline
$ \delta\varphi-\delta\chi$  &  21  & ${A}_{\alpha\beta}$ & tr$F_+^2\bar{\lambda}\bar{\lambda}$   &7\\
\hline
\end{tabular}

\end{center}

We see that
the dilaton $\phi$ is a flat direction. The negative mass-square is of course above the Breitenlohner-Freedman bound (\ref{bf}).

In the third column of the table above we have indicated the ten-dimensional origin of the
corresponding perturbation, in the notation of \cite{kim}. Namely,  $B$ is related to the axion-dilaton; $h_{\alpha}^\alpha$ and ${h}_{(\alpha\beta)}$ are the trace and traceless part respectively of the internal-space metric; ${A}_{\alpha\beta}$ is the internal part of the complex two-form potential.
In the fourth column ($\mathcal{O}$) we have listed the dual CFT$_4$ operators,
in the notation used in table 7 of \cite{dhoker}, with their corresponding
dimensions ($\Delta$) given in the last column.

The table 7 of \cite{dhoker} also indicates how operators are obtained as
supersymmetric descendants of the
operators\footnote{Strictly speaking, our case is the degenerate case, $k=0$.} tr$X^k$.
This implies that the spectrum of operators listed above has the following structure:

\begin{center}
\begin{tabular}{| c ||c|c|}
\hline
$\Delta$ & $\mathcal{O}$ & $Q$ descendant \\
\hline
\hline
3 & $\mathcal{O}_3 :=$ tr$\lambda\lambda$  & $Q^2$ \\
\hline
4 & $\mathcal{O}_4 :=$ tr$F_+^2$  & $Q^4$ \\
\hline
6 & $\mathcal{O}_6 :=$ tr$\lambda\lambda\bar{\lambda}\bar{\lambda}$  & $Q^2 \bar{Q}^2$ \\
\hline
7 & $\mathcal{O}_7 :=$ tr$F_+^2\bar{\lambda}\bar{\lambda}$  &   $Q^4 \bar{Q}^2$ \\
\hline
8 & $\mathcal{O}_8 :=$  tr$F_+^2 F_-^2$ & $Q^4 \bar{Q}^4$ \\
\hline
\end{tabular}
\end{center}

From this table we see that the only non-vanishing terms in the operator product expansions are:
\eq{
\mathcal{O}_3 \mathcal{O}_6 \sim \mathcal{O}_7~;~~~~~
\mathcal{O}_3 \mathcal{O}_3 \sim \mathcal{O}_4 ~;~~~~~
\mathcal{O}_6 \mathcal{O}_6 \sim \mathcal{O}_8
~,
}
with all other products vanishing because they contain more than four $Q$ or $\bar{Q}$. This argument from the
field theory justifies why this set of five operators has a closed OPE, and thus explains why there is a consistent supergravity truncation to these modes.

In \cite{italians} a related consistent truncation of supergravity was obtained which contains three additional scalar fields. The additional modes are dual to operators $\mathcal{O}_3' :=$ tr$\bar{\lambda} \bar{\lambda} \sim \bar{Q}^2$;
$\mathcal{O}_4' :=$ tr$F_-^2 \sim \bar{Q}^4$ and $\mathcal{O}_7' :=$ tr$F_-^2 \bar{\lambda} \bar{\lambda} \sim Q^2 \bar{Q}^4$. These three operators have a trivial OPE with each other, but together with the other five operators
generate a closed OPE in which:
\eq{\spl{
\mathcal{O}'_3 \mathcal{O}_3 & \sim \mathcal{O}_6~;~~~~~
\mathcal{O}'_3 \mathcal{O}_4 \sim \mathcal{O}_7 ~;~~~~~
\mathcal{O}'_3 \mathcal{O}_6 \sim \mathcal{O}_7'~;~~~~~
\mathcal{O}'_3 \mathcal{O}_7 \sim \mathcal{O}_8~, \\
\mathcal{O}_4' \mathcal{O}_3 & \sim \mathcal{O}_7'~;~~~~~
\mathcal{O}_4' \mathcal{O}_4 \sim \mathcal{O}_8~;~~~~~
\mathcal{O}_7' \mathcal{O}_3 \sim \mathcal{O}_8~.
}}
The existence of this closed OPE of eight operators thus explains the origin of the consistent supergravity truncation in this case also.

\subsection{The non-supersymmetric $AdS_5\times \mathcal{M}_5$}

We expand the fields around the critical point as follows:
\eq{\spl{
\phi&=\phi_0+\delta \phi\\
e^{2\psi}&=\frac{2}{\sqrt{5}}(1+\delta\psi) \\
e^{2\omega}&=\frac{\sqrt{5}}{2}(1+\delta\omega)  \\
{\varphi}&=\sqrt{2}e^{-\phi_0/2}\left( 1+ \delta\varphi \right)\\
{\chi}&= \sqrt{2}e^{\phi_0/2}\left(1+ \delta\chi \right)
~,}}
where $\phi_0$ is a constant.  The inverse radius $\lambda$ of $AdS_5$ is related to the constant $W$ through eq.~(\ref{61}).
%
%

Expanding the equations of motion
to first order in the fluctuations we find the following mass spectrum:

\bigskip

\begin{center}

\begin{tabular}{| c || c | c |}
\hline
$\sigma$ &  $(M_\sigma/\lambda)^2$ & $\Delta$ \\
\hline
\hline
$2\delta \phi - \delta\varphi+\delta\chi$  &  0  & 4  \\
\hline
$-\delta \phi - \delta\varphi+\delta\chi$ &  24 & 2 (1 + $\sqrt{7}$)  \\
\hline
$\delta\psi+\delta\omega$  &  32 & 8 \\
\hline
$4\delta\psi+ \delta\varphi+\delta\chi$ &    32 & 8  \\
\hline
$-2\delta\psi+ \delta\varphi+\delta\chi$  & 8 &  2 (1 + $\sqrt{3}$) \\
\hline
\end{tabular}

\end{center}

As in the supersymmetric solution, there is again a single flat direction and all mass-squares are positive.
Note that an operator of dimension $2(1+ \sqrt{3})$ was also found in the analysis of the spectrum
around the non-supersymmetric $AdS_5$ vacuum in \cite{Distler:1998gb}.

\section{Conclusions}\label{conc}

One of the motivations for this work was to develop holography for $AdS_4 \times \mathbb{R} \times \mathcal{M}_5$ IIB solutions. In this paper we have shown that both supersymmetric and non-supersymmetric solutions can be obtained
as solutions of five-dimensional consistent truncations. As a next step one would like to use the truncated action
as a starting point for setting up a holographic dictionary. Recall that in the familiar case of $AdS_5$, the
consistent truncation to five-dimensional Einstein
gravity with cosmological constant captures only the stress energy tensor of the dual conformal theory. By considering asymptotically locally $AdS_5$ solutions of the form
\be
ds^2 = \frac{d \rho^2}{4 \rho^2} + \frac{1}{\rho } \left ( g_{(0) ij} dx^i dx^j + \rho g_{(2) ij} + \rho^2 g_{(4) ij} \cdots \right ) dx^i dx^j,
\ee
near the AdS conformal boundary as $\rho \rightarrow 0$, in which $g_{(0) ij}$ acts as the source for the CFT stress energy tensor, we set up a holographic dictionary by analyzing such geometries and computing renormalized correlators of the CFT stress energy tensor, see \cite{de Haro:2000xn}. An important step in this analysis is to generalize the $AdS_5$ solution to solutions which asymptotically locally approach $AdS_5$.

Returning now to the case of $AdS_4 \times \mathbb{R}$, the key issue is to establish the structure of the conformal boundary. Note that since the coordinate along ${\mathbb{R}}$ is non-compact one cannot dimensionally reduce to four dimensions; more precisely, the conformal boundary of
the five-dimensional spacetime is not the conformal boundary of $AdS_4$. A related discussion
appeared in the context of the Janus solution. These solutions are $AdS$ sliced domain walls, in which the conformal boundary consists of two half-spheres with angular excess joined at the equator, and the coupling constant (dilaton) takes different values on either side; see for example section 4.1 of \cite{ps}. In fact the non-supersymmetric $AdS_4 \times \mathbb{R}$ can be obtained as a particular limit of the Janus solution; however, the analysis of \cite{ps} degenerates in this limit.

One natural idea is to change the conformal frame in five dimensions to obtain a metric which is asymptotically locally $AdS_5$. This idea was previously applied to non-conformal branes
(i.e. running dilaton backgrounds, see \cite{precision}).
The non-conformal brane geometries admit a distinguished conformal frame in which the geometry becomes $AdS$ with the dilaton running; the geometries capture the dual theory in the region where the dynamics is driven by the dimensionality of the coupling constant, see for example section 3 of \cite{precision}. The $AdS_4 \times \mathbb{R}$  solutions also become asymptotically locally AdS by a change of frame and it would be interesting to exploit this fact in order to set up the holographic dictionary for this class of solutions. We hope to return to this issue in the future.

It seems possible that the supersymmetric Janus solutions of \cite{janus} may also be recovered within the framework of our effective action.
The authors of that reference constructed their solutions using (a deformation
of) $S^5$ as their internal space. Since our reduction ansatz  makes no reference to any specific $\mathcal{M}_5$, as long as it admits
a Sasaki-Einstein structure,
if the Janus solutions can indeed be obtained using
our effective action
 this would demonstrate that they admit straightforward generalizations to arbitrary five-manifolds admitting Sasaki-Einstein  structure. It would be interesting to explore this further.

{\bf Note added:}
While this paper was being prepared, we received preprint
\cite{italians} which has significant overlap with our results. In particular,
it should be possible to show that the five-dimensional effective action (\ref{leff}) of the present paper is a truncation to the
graviton coupled to five real scalars of the supersymmetric five-dimensional effective action presented in \cite{italians}.
The mass fluctuations around the two $AdS_5$ vacua presented in our section
\ref{masses} can
also be identified with the corresponding mass spectra presented in tables
(5.6) and (5.16) of \cite{italians}.\footnote{Note, however, that there is an overall factor of 8/9 discrepancy in the mass spectra around the non-supersymmetric  $AdS_5$ vacuum between our results and those of \cite{italians}; our
masses are normalized with respect to the inverse radius of
the corresponding $AdS_5$ space.  Moreover  our results suggest that
there is a typo in the fifth line of table (5.16) of \cite{italians}
and one should replace
$(\mathrm{Im}\delta b^\Omega,\mathrm{Re}\delta c^\Omega)$ with
$(\mathrm{Re}\delta b^\Omega,\mathrm{Im}\delta c^\Omega)$.}

\section*{Acknowledgments}

M.T. and D.T. would like to thank  the organizers of the 5th Regional Meeting in String Theory in Kolymbari, Crete, where this work was initiated. This work is part of the research program of the Stichting voor Fundamenteel Onderzoek
der Materie (FOM), which is financially supported by the Nederlandse Organisatie
voor Wetenschappelijk Onderzoek (NWO). Two of the authors acknowledge support
from NWO: K.S. via a Vici grant, M.T. via the Vidi grant 'Holography, duality and time
dependence in string theory'.

\appendix

\section{Five-dimensional Sasaki-Einstein manifolds}\label{app:se}

As already mentioned in section \ref{sec1}, the internal
five-dimensional manifold $\mathcal{M}_5$ of our ten-dimensional
reduction ansatz can be obtained from a Sasaki-Einstein manifold
through warping and squashing. Furthermore, associated with
the Sasaki-Einstein structure there is an underlying
$SU(2)$ structure which
was crucial for obtaining
the supersymmetric solutions of \cite{lt},
and is also
central to the consistent truncation of
the present paper. In this appendix we explain some
of the relevant geometrical concepts. For further
details the reader may consult \cite{lt}.

In five dimensions a Sasaki-Einstein manifold may be defined under certain
additional mild assumptions
 as one which admits a pair of Killing spinors (related to each other
by complex conjugation).
From that it follows that the metric is Einstein. With the
canonical normalization of a `unit' Sasaki-Einstein
space in five dimensions, the Ricci tensor of $\mathcal{M}_5$ is given by:
\eq{
R^{(SE)}_{mn}=4 g^{(SE)}_{mn}
~,}
so that the six-dimensional cone $\mathcal{C}(\mathcal{M}_5)$ is Calabi-Yau.

Moreover it  can be shown that one can construct
three real two-forms $\alpha$, $\beta$, $\gamma$ and a real
one-form $u$, as bilinears of the Killing spinor, obeying
the algebraic conditions:
\eq{\spl{\label{su2a}
\iota_{u}\alpha&=\iota_{u}\beta=\iota_{u}\gamma=0\\
\alpha\wedge\beta&=\beta\wedge\gamma=\gamma\wedge\alpha=0\\
\alpha\wedge\alpha&=\beta\wedge\beta=\gamma\wedge\gamma\neq0~
~.}}
Therefore, the forms $(\alpha,\beta,\gamma,u)$ define an associated
local $SU(2)$ structure. In addition the Killing spinor property can
be used to show that the following differential conditions are obeyed:
\eq{\label{su2b}
\d u=-2\gamma~;~~~~~
\d(\alpha+i\beta)=-3iu\wedge(\alpha+i\beta)~;~~~~~
\d\gamma=0
~.}

The Sasaki-Einstein metric associated with the $SU(2)$ structure (\ref{su2a},\ref{su2b}) can locally
 be put in the canonical form:
\eq{\label{sem}
ds_{SE}^2=ds^2_{KE}+u\otimes u~,
}
where $ds^2_{KE}$ is a K\"{a}hler-Einstein four-dimensional base over which $u$ is fibered.
The connection field strength of this local $U(1)$ fibration is the K\"{a}hler form of the base,
and is equal to $du=-2\gamma$. If in addition the orbits of the vector  dual to $u$ (which is known as the `Reeb vector')
are closed and the associated $U(1)$ action is free, (\ref{sem}) extends globally and the base is a four-dimensional
K\"{a}hler-Einstein manifold of positive curvature.

In appendix \ref{app:truncation} we explain in detail
how the local $SU(2)$ structure $(\alpha,\beta,\gamma,u)$ can be used to build the reduction ansatz leading to the consistent truncation.

\section{Consistent truncation}\label{app:truncation}

In this appendix we give some technical details
concerning the consistent truncation presented
in section \ref{sec1}.

Our starting point is  the
ten-dimensional type  IIB supergravity in the string frame. The equations
of motion can be derived from the following pseudoaction:
\eq{\label{pseudo}
S=\frac{1}{2\kappa^2_{10}}\int\d^{10}x\sqrt{-g}\Big\{e^{-2\phi} \big[R+4(\partial\phi)^2-\frac12 H^2] -\frac14 F^2 \Big\}~,
}
where we are using polyform notation.
Our choice of
notation and conventions follows appendix A of \cite{lmmt},
which the reader may consult for further details.

Let us now describe the reduction ansatz. For the 10D
metric in the string frame we set:
\eq{\label{g10}
ds^2_{10}=e^{2A(x)}\left\{g_{\mu\nu}dx^\mu dx^\nu + ds^2(\mathcal{M}_5) \right\}
~,}
where ($\mu$,$\nu$) are indices along the 5D spacetime, and $\mathcal{M}_5$ is the five-dimensional internal manifold.
Furthermore we shall take the
internal metric to be of the form:
\be\label{5dmetric}
ds^2(\mathcal{M}_5)=c_1~\! e^{2\psi(x)}ds^2_{KE}
+c_2~\!e^{2\omega(x)}u\otimes u~,
\ee
where
\be
ds^2_{KE}+ u\otimes u~,
\ee
is the metric of a canonically normalized `unit' five-dimensional
Sasaki-Einstein space ({\it cf.} appendix \ref{app:se}). All three functions $A$, $\psi$, $\omega$ are assumed to depend only
on the 5D spacetime coordinates $x^{\mu}$. The constants
$c_1$, $c_2$ in (\ref{5dmetric}) are chosen as follows:
\eq{\label{25}
c_1=\frac{6}{5W^2}~,~~~~~c_2=\left(\frac{6}{5W}\right)^2~,
}
so that in the $\psi=\omega=0$ limit the internal metric (\ref{5dmetric}) reduces to the squashed
Sasaki-Einstein internal space of the type IIB $\mathcal{N}=1$
solution of \cite{lt}. $W$ is an arbitrary real constant within
supergravity, however upon imposing flux quantization it
will be constrained to discrete values.

For the dilaton we assume:
\be\label{phiansatz}
\phi=4A~,
\ee
so that the 10D Einstein-frame metric (\ref{g10e}) is obtained from (\ref{g10}) by simply dropping the overall warping.

The ansatz for the NSNS
three-form flux is given by:
\be\label{nsflux}
H= \frac{1}{4f}W \mathrm{Re}\Omega\wedge d\chi
-2\sqrt{c_2}f {\chi}J\wedge u
~,
\ee
where  $\chi(x)$ is a 5D spacetime scalar, and we have set:
\bea\label{30}
f := \frac{\sqrt{5}}{4}W~,
\eea
for later convenience;
the  two-forms $J$, $\Omega$ are related to the underlying
$SU(2)$ structure of $\mathcal{M}_5$ via:
\bea
J&=&c_1 \left(\sin\theta~\!\alpha+\cos\theta~\!\beta\right)\nn\\
\Omega&=&c_1 \left(\cos\theta~\!\alpha-\sin\theta~\!\beta-i\gamma\right)~,
\eea
where $\theta$ is constant. The real two-forms
$\alpha$, $\beta$, $\gamma$ and the real one-form $u$
define a local $SU(2)$ structure on the internal space,
obeying the algebraic and
differential conditions given in eqs.~(\ref{su2a},\ref{su2b}).

The RR fluxes are given by:
\bea\label{rrflux}
e^{\phi}F_1&=&0\nn\\
e^{\phi}F_3&=&\frac{1}{4f}W J\wedge d\varphi
+2\sqrt{c_2}f {\varphi}\mathrm{Re}\Omega\wedge u\nn\\
e^{\phi}F_5&=&(2-\frac{1}{2}{\chi\varphi})\sqrt{c_2}W J\wedge J\wedge u
~.
\eea
The total five-form flux is self-dual
and given by $F_5^{tot}=F_5+\star_{10} F_5$.

The above ansatz guarantees that the fluxes obey the Bianchi identities:
\eq{
dH=0~,~~~~~dF+H\wedge F=0
~,}
where we are using polyform notation.

 The combination
$R^{(10)}_{MN}+2\nabla^{(10)}_M\partial_N\phi$, where $R^{(10)}_{MN}$ is the Ricci tensor and $\nabla^{(10)}_M$ the covariant derivative associated with the metric in (\ref{g10}), appears in the ten-dimensional
string-frame Einstein equations; ($M$,$N$) are
ten-dimensional Einstein indices. Taking (\ref{g10},\ref{phiansatz}) into account, we compute:
\eq{
\spl{\label{useful}
(R^{(10)}&+2\nabla^{(10)}\partial\phi)_{\mu n}=0\\
(R^{(10)}&+2\nabla^{(10)}\partial\phi)_{\mu\nu}
=R^{(5)}_{\mu\nu}-g_{\mu\nu}\left(\nabla^2A+\partial(4\psi+\omega)\cdot\partial A\right)\\
&~~~~~~~~~~~~~~~~~~~~~-8\partial_\mu A\partial_\nu A
-4\partial_\mu \psi\partial_\nu \psi-\partial_\mu \omega\partial_\nu \omega
-\nabla_\mu \partial_\nu( 4 \psi +\omega)
\\
(R^{(10)}&+2\nabla^{(10)}\partial\phi)_{mn}=\\
&-u_m u_n ~\! c_2e^{2\omega}\left\{
\nabla^2(A+\omega)+\partial(4\psi+\omega)\cdot\partial(A+\omega)-
\frac{4c_2e^{2\omega}}{c_1^2e^{4\psi}}
\right\} \\
&-\tilde{g}_{mn}\left\{
\nabla^2(A+\psi)+\partial(4\psi+\omega)\cdot\partial(A+\psi)-
\frac{6}{c_1e^{2\psi}}\left(1-\frac{c_2e^{2\omega}}{3c_1e^{2\psi}}
\right)\right\}
~,}
}
where the metric $\tilde{g}_{mn}$ is given by:
\be\label{5dricci}
\tilde{g}_{mn}dx^m\otimes dx^n=c_1 e^{2\psi}ds^2_{KE}~.
\ee

In addition to the above, in order to reduce the 10D equations of motion
using our ansatz, one needs to make use of the identities obeyed by the $SU(2)$ structure  listed in \cite{blt, ltscalar}. Taking into account all of the
above, we finally obtain the following:

\begin{itemize}

\item The 10D dilaton eom reduces to
\eq{\spl{\label{r1}
\nabla^{\mu}\left(e^{4\psi+\omega}\partial_\mu A\right)
&=\frac{W^2}{64f^2}\left\{
e^{4A+\omega}(\partial\varphi)^2
-e^{-4A+\omega}(\partial\chi)^2
\right\}\\
&+f^2\left(
e^{4A-\omega}\varphi^2
-e^{-4A-\omega}\chi^2
\right)
~.}}
\item The internal piece of the 10D Einstein eom reduces to
two scalar equations
\eq{\spl{\label{r2}
\nabla^{\mu}\left(e^{4\psi+\omega}\partial_\mu \psi\right)
&=-\frac{W^2}{64f^2}\left\{
e^{4A+\omega}(\partial\varphi)^2
+e^{-4A+\omega}(\partial\chi)^2
\right\}\\
&-W^2(2-\frac{1}{2}{\chi\varphi})^2e^{-4\psi-\omega}+
\frac{6}{c_1}\left(1-\frac{c_2e^{2\omega}}{3c_1e^{2\psi}}
\right)e^{2\psi+\omega}
\\
&-f^2\left(
e^{4A-\omega}\varphi^2
+e^{-4A-\omega}\chi^2
\right)
~}}
and
\eq{\spl{\label{r3}
\nabla^{\mu}\left(e^{4\psi+\omega}\partial_\mu \omega\right)
&=\frac{W^2}{64f^2}\left\{
e^{4A+\omega}(\partial\varphi)^2
+e^{-4A+\omega}(\partial\chi)^2
\right\}\\
&-W^2(2-\frac{1}{2}{\chi\varphi})^2e^{-4\psi-\omega}+
\frac{4c_2}{c_1^{2}} e^{3\omega}\\
&-3f^2\left(
e^{4A-\omega}\varphi^2
+e^{-4A-\omega}\chi^2
\right)
~.}}

\item The mixed piece of the 10D Einstein eom is
automatically satisfied.

\item The external piece of the 10D Einstein eom reduces to
\eq{\spl{\label{r4}
0&=e^{4\psi+\omega}R^{(5)}_{\mu\nu}
+e^{4\psi+\omega}\left(
12\partial_\mu\psi\partial_\nu\psi+8\partial_\mu\psi\partial_\nu\omega
-8\partial_\mu A\partial_\nu A\right)
\\
&-\frac{W^2}{16f^2}\left(
e^{4A+\omega}\partial_\mu\varphi\partial_\nu\varphi
+e^{-4A+\omega}\partial_\mu\chi\partial_\nu\chi\right)
-\nabla_\mu\partial_{\nu}e^{4\psi+\omega}\\
&-\frac{1}{3}g_{\mu\nu}\Big\{
\nabla^2e^{4\psi+\omega}
+2{W^2}(2-\frac{1}{2}{\chi\varphi})^2e^{-4\psi-\omega}\\
&~~~~~~~~~~-
\frac{24}{c_1}\left(1-\frac{c_2e^{2\omega}}{6c_1e^{2\psi}}
\right)e^{2\psi+\omega}+4f^2\left(
e^{4A-\omega}\varphi^2
+e^{-4A-\omega}\chi^2
\right)
\Big\}
~,}}
where we have taken (\ref{r1}) into account.

\item The $H$-form eom  reduces to
\eq{\spl{\label{r5}
\nabla^{\mu}\left(e^{-4A+\omega}\partial_\mu \chi\right)
=20f^2e^{-4A-\omega}\chi
-16f^2\varphi(2-\frac{1}{2}{\chi\varphi}) e^{-4\psi-\omega}
~.}}

\item The $F_1$-form eom is
automatically satisfied.

\item The $F_3$-form eom reduces to
\eq{\spl{\label{r6}
\nabla^{\mu}\left(e^{4A+\omega}\partial_\mu \varphi\right)
=20f^2e^{4A-\omega}\varphi
-16f^2\chi(2-\frac{1}{2}{\chi\varphi}) e^{-4\psi-\omega}
~.}}

\item The $F_5$-form eom  is
automatically satisfied.

\end{itemize}

It is now straightforward to see that equations (\ref{r1}-\ref{r6}) can be `integrated'
to the five-dimensional effective action given in eq.~(\ref{leff}) of section \ref{sec1}. As might have been expected, the Lagrangian (\ref{leff}) turns out to be equivalent to the string-frame ten-dimensional IIB pseudoaction in
eq.~(\ref{pseudo})
upon substitution of the ansatz (\ref{g10},\ref{phiansatz},\ref{nsflux},\ref{rrflux}) and
integration over the internal directions.

\end{document}